\documentclass[aps,prl,reprint,superscriptaddress,amsmath,amssymb,twocolumn]{revtex4-1}
\usepackage{graphicx,scrtime}
\newcommand{\be}{\begin{equation}}
\newcommand{\ee}{\end{equation}}
\newcommand{\im}{\mbox{Im}}
\newcommand{\ba}{\begin{eqnarray}}
\newcommand{\ea}{\end{eqnarray}}

\newcommand{\rr}{{\bf r}}
\newcommand{\qq}{{\bf q}}
\begin{document}
{\sf Version v7, \today, \thistime}
\title{Absence of a boson peak in anharmonic phonon models with Akhiezer-type damping}

\author{Andrij Shvaika}
\affiliation{Institute for Condensed Matter Physics of the National Academy of Sciences of Ukraine, UA-79011 Lviv, Ukraine}

\author{Mykola Shpot}
\affiliation{Institute for Condensed Matter Physics of the National Academy of Sciences of Ukraine, UA-79011 Lviv, Ukraine}

\author{Walter Schirmacher}
\affiliation{Institut f\"ur Physik, Universit\"at Mainz, D-55099 Mainz, Germany}

\author{Taras Bryk}
\affiliation{Institute for Condensed Matter Physics of the National Academy of Sciences of Ukraine, UA-79011 Lviv, Ukraine}
\affiliation{Institute of Applied Mathematics and Fundamental Sciences, Lviv National Polytechnic University, UA-79013 Lviv, Ukraine}

\author{Giancarlo Ruocco}
\affiliation{Center for Life Nano Science @Sapienza, Istituto Italiano di Tecnologia, 295 Viale Regina Elena, I-00161, Roma, Italy}
\affiliation{Dipartimento di Fisica, Universita' di Roma ``La Sapienza'', I-00185, Roma, Italy}
\begin{abstract}
	In a recent article M. Baggioli and A. Zaccone (Phys. Rev. Lett. {\bf 112}, 145501 (2019)) claimed that an anharmonic damping, leading to a sound
	attenuation proportional to $\omega^2$ (Akhiezer-type damping)
	would imply a boson peak, i.e.\ a maximum in the vibrational density
	of states, divided by the frequency squared (reduced
	density of states). This would apply
	both to glasses and crystals.
	Here we show that
	this is not the case. In a mathematically correct
treatment of the model
	the reduced density of states monotonously decreases, i.e.\
	there is no boson peak.
	We further show that the formula for the would-be boson peak,
	presented by the authors, corresponds to a very short one-dimensional
	damped oscillator system. The peaks they show
correspond to
	resonances,
	which vanish in the thermodynamic limit.
\end{abstract}
\date{\today}

\maketitle

One of the characteristic features of disordered solids, in particular,
of glasses, is an anomalous excess of the vibrational density of states
(DOS) $g(\omega)$ beyond the Debye law $g_D(\omega)
\propto\omega^2$, which appears
as a peak in the reduced density of states $g(\omega)/\omega^2$.
This peak has been called ``boson peak'' (BP) in the early literature
on the spectral properties of glasses \cite{jackle81}, because
the temperature dependence of the measured Raman spectra followed
the boson occupation function $n(\omega)+1=1/[1-\exp\{-\hbar\omega/k_BT\}]$,
pointing to a temperature-independent spectrum. As all anharmonic phenomena
are temperature dependent, the conclusion was, that the BP in
glasses must be a harmonic phenomenon.

A vast number of suggestions for the origin of the BP has been
published in the last 60 years
(see e.g.\ \cite{klinger10,schirm14} for bibliographies). Most authors
attribute the anomaly to the frozen-in structural disorder, be it
via spatially fluctuating elastic constants \cite{schirm06,schirm07,marruzzo13,schirm14} or quasi-localized soft-potential defects \cite{karpov83,buchenau91}.
Some other authors interpret the BP as a
crystal-like van-Hove singularity, washed out by the structural disorder
\cite{te01,zorn11,chumakov11,chumakov14}.
All three mechanisms nowadays are known to exist separately
\cite{kapteijns18,wang18,kapteijns21}, and are, as said above,
of harmonic origin.

In spite of the established evidence for the harmonic origin
of the BP, in a recent paper
Baggioli and Zaccone (BZ) \cite{baggioli19} attributed it
to the anharmonicity-caused damping of acoustic modes.
Guided by the observation of boson-peak-like features in
the temperature dependence of the specific heat of some
crystals
\cite{chumakov14,moratalla15,szewczyk15,jezowski18}, they claimed
that the BP anomaly in glasses and crystals would
be universally of anharmonic origin.

However, their treatment contains severe mathematical errors. We show
in the remainder of this article that a correct treatment of their
anharmonic phonon model does not produce a maximum in the
reduced DOS $g(\omega)/\omega^2$, i.e.\ no BP,
thus rendering all their conclusions irrelevant and useless
for discussing the BP vibrational anomalies. In fact, as we shall
show, the ``boson peaks'' they presented
correspond to resonances of a
one-dimensional system. These resonances
only exist for very short system lengths and go away in
the thermodynamic limit.

BZ consider anharmonically damped longitudinal \mbox{($\alpha\!=\! L$)}
and transverse \mbox{($\alpha\!=\! T$)} phonons, obeying hydrodynamic equations
\be\label{e1}
\frac{\partial^2}{\partial t^2}u_\alpha(\rr,t)
=\nabla^2\bigg(c_\alpha^2+D_\alpha\frac{\partial}{\partial t}
\bigg)u_\alpha(\rr,t)
\ee
with $\alpha=(L,T)$, $c_\alpha$ the sound velocities and $D_\alpha$
the damping coefficients. Eq.~(\ref{e1}) reads
in wavenumber and frequency space
\be\label{e2}
\Big[\omega^2-q^2\left(c_\alpha^2-i\omega D_\alpha\right)\Big]u_\alpha(\qq,\omega)=0\,.
\ee
A damping term $\propto i\omega$ corresponds to sound attenuation
according to
$\Gamma(\omega)\propto \omega^2$, which has been derived for
crystals from an anharmonic Hamiltonian by Akhiezer \cite{akhiezer39} and
for amorphous solids by Tomaras et al.~\cite{tomaras10}. Both theories
give a linear temperature increase of the coefficients $D_\alpha$.

It has been pointed out by some of the present authors \cite{schirm07}
that in the frequency range near and below the boson peak a relationship
between the excess over the Debye density of states and the sound
attenuation excists, i.e.\ $\Delta g(\omega)=g(\omega)-g_D(\omega)
\propto \Gamma(\omega)$. Therefore, for a sound attenuation
proportional to $\omega^2$ in addition to the Debye $\omega^2$ law,
a second $\omega^2$ contribution can be expected, if Akhiezer-type
anharmonic is efficient. Because
this contribution, divided by $\omega^2$ is
constant, i.e.\ does not increase
with frequency, it does not lead to a BP. This will be demonstrated
below in detail.

However, BZ claim to be able to produce boson peaks, i.e.\ peaks
in the reduced DOS $g(\omega)/\omega^2$ from their anharmonic model. They
proceed by relating the vibrational DOS in the usual way to the Green's
functions. In three dimensions ($d$ = 3) one has
\ba\label{e3}
g(\omega)&=&\frac{1}{3}\bigg(g_L(\omega)+2g_T(\omega)\bigg)\nonumber\\
&=&-\frac{1}{3}\frac{2\omega}{\pi}\im\bigg\{
	G_L(\omega)+2G_T(\omega)
	\bigg\}
\ea
with the local Green's functions
\be\label{e4}
G_\alpha(\omega)=\frac{1}{N}\sum_\qq
\frac{1}{\omega^2-c_\alpha^2(\omega)q^2}\, ,
\ee
where $c_\alpha^2(\omega)=c_\alpha^2-i\omega D_\alpha$.

The standard procedure for the
{\bf q} sum is to consider a cubic sample of size $L$ with
periodic boundary conditions. This gives a triple sum over discrete components
$q_i=\nu_i\Delta q$, $i=x,y,z$ with $\nu_i\in\mathbb{Z}$ and $\Delta q=2\pi/L$.
In the limit $L\rightarrow\infty$, it transforms to the standard integral via
$N^{-1}\sum_{\bf q}\rightarrow N^{-1}(L/2\pi)^3
\int \mathrm{d}q_x
\int \mathrm{d}q_y
\int \mathrm{d}q_z
=3q_D^{-3}\int_0^{q_D}\mathrm{d}q\, q^2$ with the Debye wavenumber
$q_D=(6\pi^2N/L^3)^{1/3}$, $N$ being the number of atoms.

In their
paper BZ say that the $q$ integral for calculating $G_\alpha(\omega)$
would not be analytical. The integral from
$q=0$ to $q=q_D$
is elementary, and is given by
\be\label{e5}
	G_{\alpha}^{3d}(\omega)
	=-\frac{3}{\xi_\alpha^2(\omega)} \bigg(1+\frac{\eta_\alpha(\omega)}{2}\ln\frac{\eta_\alpha(\omega)-1}{\eta_\alpha(\omega)+1}\bigg)
\ee
with $\xi_\alpha(\omega)=q_D c_\alpha(\omega)$ and
$\eta_\alpha(\omega)=\omega/\xi_\alpha(\omega)$.

\begin{figure}
	\includegraphics[width=.35\textwidth]{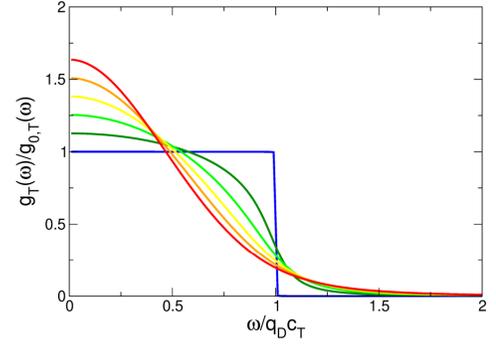}
	\caption{(Color online) Reduced transverse density of states
	$g_T(\omega)/g_{0,T}(\omega)$ vs.
	$\omega/q_Dc_T$ for $D_Tq_D/c_T$ = 0.0, 0.2, 0.4, 0.6, 0.8, 1.0 (from blue to red curves, respectively).
	}\label{fig1}
\end{figure}

The function $G(\eta_\alpha)$ is analytical in the entire complex plane,
except on the real axis. Moreover, this exact result does not give
any maximum in the reduced density of states. This can be seen
from Fig.~\ref{fig1}, where we have plotted the transverse
reduced DOS $g_T(\omega)/g_{0,T}(\omega)$ with
\be
g_{0,T}(\omega)=\frac{3}{q_D^3c_T^3}\omega^2.
\ee
We see that the introduction of the anharmonic damping increases
the initial $\omega^2$ law --- as anticipated above --- and
smoothes the sharp Debye cutoff. There is no trace of a boson-peak-like
maximum. So the question arises, where do the maxima in Figs.~1 and 4
of BZ come from, which they extensively discussed as anharmonicity-induced boson peaks.

\begin{figure}
	\includegraphics[width=.35\textwidth]{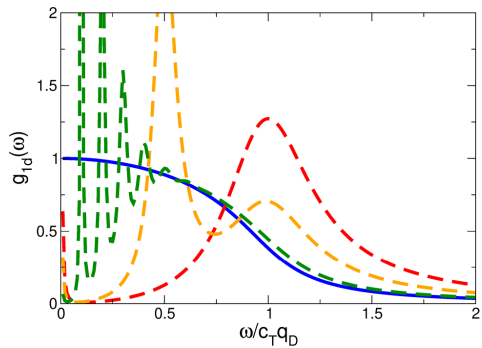}
	\caption{(Color online) One-dimensional
	transverse density of states
	$g_{T,1d}(\omega)$ vs.
	$\omega/q_Dc_T$ for $q_D/\Delta q=N/2$ = 1, 2, and 10 
	as calculated from 
    (\ref{1da})
	(dashed curves), compared with the result (\ref{1db}) for
	the limit $N\rightarrow \infty$ (blue solid curve). For the damping parameter
	we chose $D_Tq_D/c_T=0.5.$
	}\label{fig2}
\end{figure}

BZ have claimed to be able to present an alternative way to
perform the $\qq$ sum in Eq.~(\ref{e4}). The expression they
came up with
(Eq.~(7) in BZ) is rather intransparent and
involves the digamma function $\psi(z)=\mathrm{d}\ln\Gamma(z)/\mathrm{d}z$.
In a less disguised form,
identifying their variables $x,y$ with
$x=-q_D \eta_L(\omega)$, $y=q_D \eta_T(\omega)$, and using $1+i=\sqrt{2i}$,
an analogue of the Eq.~(7) of BZ can be obtained with Green's functions
\be
	\label{exact0}
	G_\alpha^{BZ}(\omega)=-
	\frac{1}{N}\frac{q_D}{2 \eta_\alpha(\omega) \xi^2_\alpha(\omega)}
	F(q_D,q_D \eta_\alpha(\omega))\, ,
\ee
where
\be\label{ffn}
F(n,z)=\psi(z)-\psi(-z)-\psi(1{+}n{+}z)+\psi(1{+}n{-}z).
\ee
Apparently, BZ used the well-known \cite[entry 4.1.5.9]{PBM1} {\it single} sum
(following from the recursion relation $\psi(z{+}1)=\psi(z)+z^{-1}$)
\begin{equation}\label{exact}
	\sum_{\nu=0}^{n}\frac{1}{\nu^2-z^2}=\frac{1}{2z}F(n,z)\, ,
\end{equation}
with  $n=q_D$, thus solving a sort of one-dimensional problem instead of the original three-dimensional one.
There are more inconsistencies in the expression (\ref{exact0}): First,
the Green's function must have the dimension of an inverse squared frequency,
which does not hold for (\ref{exact0}). Secondly, in (\ref{exact}),
$n$ is an integer and $z$ a (dimensionless) complex number.
If one wanted to apply the summation formula (\ref{exact}) to a one-dimensional
problem, the corresponding Green's function would be given, for large $N$, by
\ba\label{1da}
	G_\alpha^{1d}(\omega)&=&\frac{2}{N}\sum_{\nu=0}^{N/2}
\frac{1}{\omega^2-c^2(\omega)(\Delta q)^2\,\nu^2}
\nonumber\\
&=&-\frac{1}{2\eta_\alpha(\omega) \xi^2_\alpha(\omega)}
	F\Big(\frac{N}{2},\frac{N}{2}\eta_\alpha(\omega)\Big)\, .
\ea
For large $L$, i.e.\ small $\Delta q$, by Stirling's theorem
$\lim_{z\rightarrow\infty}\ln\Gamma(z)\rightarrow z\ln z \Rightarrow
\lim_{z\rightarrow\infty}\psi(z)\rightarrow \ln z$, which gives the
correct one-dimensional Green's function
\be\label{1db}
	G_\alpha^{1d}(\omega)\stackrel{N\rightarrow\infty}{\longrightarrow}
-\frac{1}{2\eta_\alpha(\omega) \xi^2_\alpha(\omega)}
\ln
	\frac{
		\eta_\alpha(\omega)-1
		}{
		\eta_\alpha(\omega)+1
			}
\,.
\ee
In Fig.~\ref{fig2} we have plotted the transverse DOS, which resuls
from the Green's function in Eq.~(\ref{1da}) for several values
of $q_D/\Delta q=N/2$,
along with
the result corresponding to $N\rightarrow \infty$, Eq.~(\ref{1db}).
It is clearly seen that the peaks, which occur, are the resonances
corresponding to the harmonics of a very short spring. Such resonances
have nothing to do with the boson peak in a three-dimensional solid,
discussed in the paper by BZ \cite{baggioli19}. Because in Figs.~1
and 4 of this paper only one maximum is visible, we suspect that
the authors chose $q_D/\Delta q=N/2=1$, which corresponds to a
{\it diatomic molecule}. It is clearly seen that for increasing
$N$ the resonances are shifted towards small frequencies, and
gradually the $N\rightarrow\infty$ result (\ref{1db}) is recovered. In this
limit, as in three dimensions, no ``boson peaks'' appear.

It is a coincidence, that Eq.~(\ref{1da}) for small $N$ implies 
a DOS proportional to $\omega^2$. This frequency dependence comes
from the Akhiezer law used as input and has nothing to do with
three-dimensional Debye-type wave propagation, as discussed in the
paper by BZ.

In summary, we have shown, that the ``exact'' results by BZ are not
mathematically correct and correspond to the resonances of very
small strings. The model considered by BZ, i.e.\ Debye waves with
Akhiezer-type damping does not produce a boson peak.

\subsection*{Acknowledgment}
TB was supported by \mbox{NRFU grant 2020.02/0115.}

\end{document}